\documentclass{article}
\usepackage{spconf,amsmath,graphicx}

\usepackage{url}
\usepackage{caption}
\usepackage{float}

\title{A Vocoder Based Method For Singing Voice Extraction}
\name{Pritish Chandna\textsuperscript{1}, Merlijn Blaauw\textsuperscript{1}, Jordi Bonada\textsuperscript{1}, Emilia G\'omez\textsuperscript{1,2}}
\address{\textsuperscript{1}Music Technology Group, Universitat Pompeu Fabra, Barcelona, Spain\\
\textsuperscript{2}Joint Research Centre, European Commission, Seville, Spain}

\begin{document}
%
\maketitle
\begin{abstract}
This paper presents a novel method for extracting the vocal track from a musical mixture. The musical mixture consists of a singing voice and a backing track which may comprise of various instruments. We use a convolutional network with skip and residual connections as well as dilated convolutions to estimate vocoder parameters, given the spectrogram of an input mixture. The estimated parameters are then used to synthesize the vocal track, without any interference from the backing track. We evaluate our system, through objective metrics pertinent to audio quality and interference from background sources, and via a comparative subjective evaluation. We use open-source source separation systems based on Non-negative Matrix Factorization (NMFs) and Deep Learning methods as benchmarks for our system and discuss future applications for this particular algorithm.

\end{abstract}
\begin{keywords}
Source separation, deep learning, convolutional neural networks, vocoder.
\end{keywords}
\section{Introduction}
\label{sec:intro}
Audio source separation, the process of isolating individual signals from a mixture of two or more audio signals, is a well researched topic in the field of signal processing. For vocal music, source separation can be defined as separating the lead voice from the background signal, which may comprise of a mix of various instruments. While an interesting topic in itself, the isolation of the singing voice from the backing track also serves as an intermediary step for applications such as singer identification, lyrics transcription, singing voice conversion,  karaoke remixing and other Music Information Retrieval applications . 


Much research has been done for this task over the last few decades, dominated primarily by statistical methods like principal component analysis (PCA) \cite{chan2015vocal}, independent component analysis (ICA)  \cite{comon1994independent} and non-negative matrix factorization (NMF) \cite{Durrieu2009}. While effective, these methods are generally slow and lead to artifacts and distortion in the estimation of the isolated voice signal, which can be detrimental to tasks following source separation.

In the last few years, methodologies based on Deep Learning \cite{nugraha2016multichannel,grais2016single, chandna2017monoaural} have raised the bar in terms of objective evaluation metrics related to the separation task \cite{stoller2018wave, jansson2017singing} and on processing time required. Many of these techniques focus on the magnitude component of the spectrogram of the input mixture signal and use the neural network to estimate Time Frequency (TF) masks, which are applied to the mixture spectrogram to isolate the desired signal. The waveform of the signal is synthesized by either using the phase component of the mixture spectrogram or by approximating the phase using the Griffin-Lim algorithm. Some models have recently been proposed to directly work on the waveform \cite{stoller2018wave, luo2018tasnet}. 


Our research presented in this paper is inspired by some of the earliest works in source separation, which try to re-synthesize the voice signal from a mixture of the voice signal and a noise/background signal \cite{miller1973removal, oppenheim1968homomorphic}. We propose a novel approach for the paradigm of separation, wherein we estimate vocal specific vocoder parameters from a mixture signal that can be used to synthesize a version of the vocal signal present in the mixture. We hypothesize that a neural network, as a function approximator, can be trained to estimate relevant features of the underlying voice signal from mixture signal. From these features, an approximation of the original signal can be reconstructed. Since a vocal specific synthesis approach is used, the estimated vocal track has no direct interference from the backing track. In addition, we bypass the phase synthesis/estimation problem that deters other systems and the estimated vocoder features can be used directly for applications such as lyrics extraction or voice transformation. 

We evaluate our system using standard objective metrics for source separation. However, we note that the output of the system is a synthesis of the original vocal track and therefore must be evaluated as such.


We also conducted a subjective listening test focusing on three aspects of the system that we consider to be important; the intelligibility of the synthesized signal, the isolation of the vocal track from the backing track and the overall quality of the output. We compare our proposed system to two other open source source separation systems; FASST \cite{salaun2014flexible, ozerov2012general}, based on NMFs and the deep learning based DeepConvSep \cite{chandna2017monoaural}. The source code for our model is available online\footnote{\url{https://github.com/pc2752/ss_synthesis}}, as are sound examples\footnote{\url{https://pc2752.github.io/singing_voice_sep/}}, showcasing the robustness of the model with use on a real-world example.


\section{Methodology}

\subsection{WORLD Vocoder}
The WORLD vocoder \cite{morise2016world} is a speech vocoding system, commonly used for applications such as speech synthesis, manipulation and analysis. The system decomposes a speech signal into the fundamental frequency $f0$, harmonic spectral envelope and aperiodicity envelope. It has been proved that these parameters can be used to reconstruct a high quality synthesis of speech signals, even after dimensionality reduction techniques have been applied to the parameters \cite{blaauw2017neural}.

\subsection{Network Architecture}
We use a convolutional neural network (CNN), inspired by the WaveNet\cite{van2016wavenet} architecture. WaveNet is an autoregressive convolutional network based generative model, which uses skip and residual connections with dilated causal convolutions to predict the next sample in a time series based on a fixed number of previously predicted samples. A non-autoregressive version of the WaveNet has also been used for speech denoising \cite{rethage2018wavenet} and our architecture is similar to this.

\begin{figure}[H]
\centering
\includegraphics[width=0.5\textwidth]{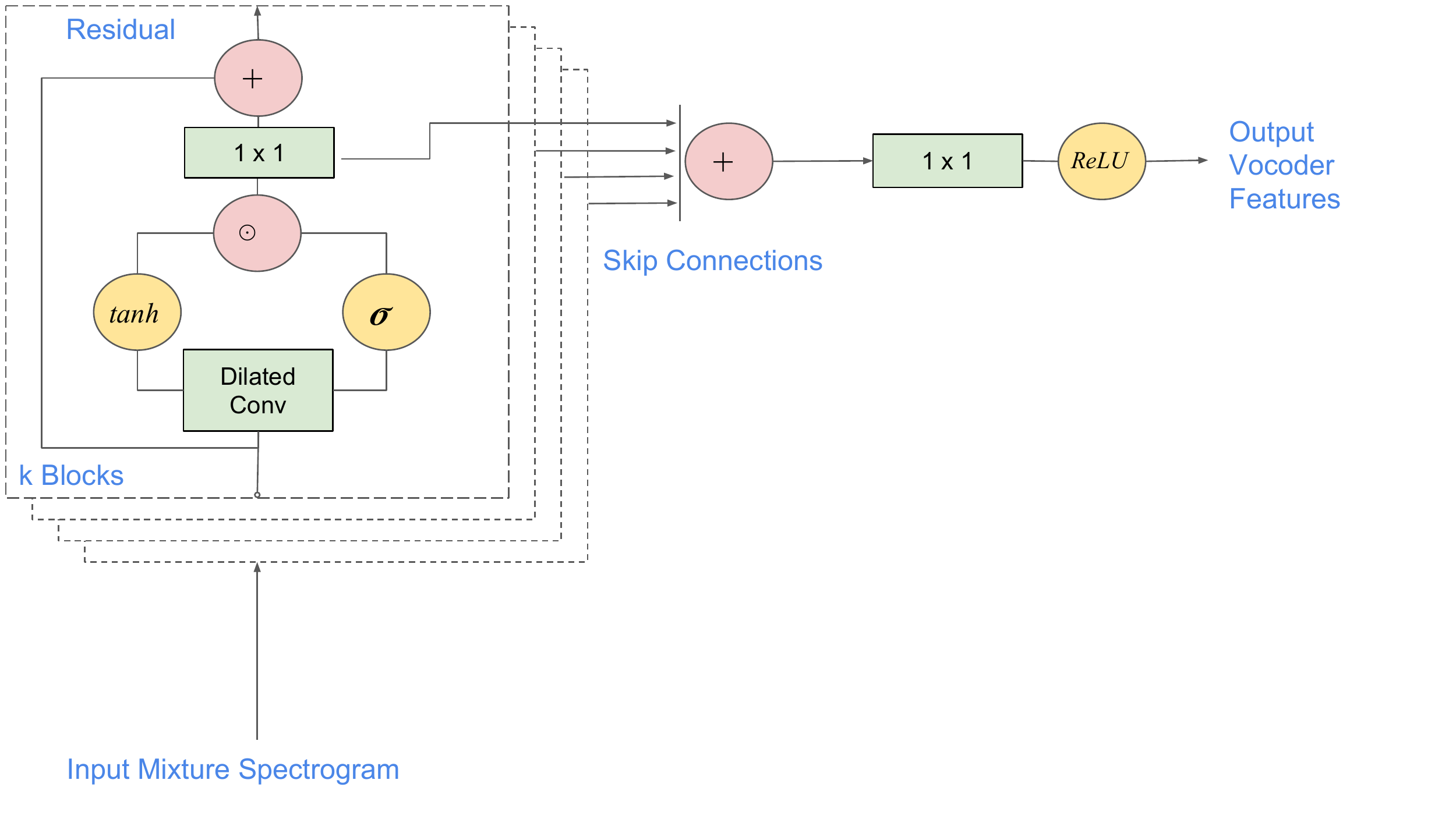}
 \caption{Overview of the convolutional block used in our model.}
 \label{fig:archi}
\end{figure}  

In our case, we use the spectrogram of the mixture signal as input and the vocoder features as the target. We use $k$ blocks of convolutional layers with skip and residual connections and gated dilated convolutions, as shown in Figure \ref{fig:archi}. To allow for fast inference, we decided to use a simpler non-autoregressive version of the architecture, similar to the one used by \cite{rethage2018wavenet}. We implement dilated convolutions, with a gated activation, as described in \cite{van2016wavenet, van2016conditional}. However, since we are not using autoregression, we do not enforce causality in the convolutional layers, but instead use zero-padding to ensure that the output of each layer has the same dimensionality in the time dimension as the previous layer.
Like \cite{blaauw2017neural}, we treat the frequency bins of the spectrogram as different channels, and thus each convolutional layer consists of one dimension convolutions across the time dimension.

The input to our network is a segment of $N$ consecutive frames of the input spectrogram. This leads to an input of dimensions $N \times D$, where $D$ is the number of bins in the spectrogram. The first two layers of the network are $1\times1$ convolutional layers. This is followed by a series of gated stacks of $2\times1$ dilated convolutions (denoted by $*$), with a sigmoid (denoted by $\sigma$) non-linearity. A similar operation is carried out with a tanh non-linearity and an element-wise multiplication (denoted by $\odot$) is applied to the result of the two, as shown in Equation \ref{eq:archi1}

\begin{equation}
z = tanh(W_{f,k}*x)\odot \sigma(W_{g,k}*x)
\label{eq:archi1}
\end{equation}

Where $W$ denotes a convolution filter and $f$ and $g$ represent filter and gates, respectively. $x$ and $z$ represent the input and output of the layer. The dilation is increased by a factor of $2$ after each block, thus exponentially increasing the time context covered through each consecutive stack. After the series of stacked convolutions, we apply two $1\times1$ convolutional layers to ensure that the output of the network has the same dimensionalty as the target vocoder features. Aside from the final layer, which has the same filter channels as the target dimension, each of the convolutional layers has $C$ channels.

\subsection{Loss function and Optimization technique}
An $L1$ loss, representing the absolute distance between the estimated parameters and the target parameters was used as the loss function for the network. Optimization was done using the Adam \cite{kingma2014adam} optimizer.

\section{EXPERIMENTS}

\subsection{Dataset and Pre-processing}
The iKala dataset \cite{chan2015vocal} was used for our experiments. This dataset, contains $252$, $30$ second tracks of vocal and backing track music as well as manually annotated MIDI-note pitch annotations for each of the vocal tracks. A cappella vocal tracks from multiple male and female singers are present in this dataset without external effects such as reverb or compression, thus making it ideal for our experiments as they hinder the performance of the vocoder. We split the dataset into a training set of $226$ songs for training the model and a test set of $26$ songs for testing the output. 

For these songs, the Short Time Fourier Transform (STFT), WORLD parameters and fundamental frequency ware calculated using Fourier transform of $1024$ bins, a hop time of $5$ms and a sampling rate of $44.1$ kHz. This led to a spectrogram with $D=513$ frequency bin and harmonic and aperiodic vocoder features with $1024$ bins per time frame. Like  \cite{blaauw2017neural}, we reduce dimensionality of the harmonic components using truncated frequency warping in the cepstral domain \cite{tokuda1994mel} with an all-pole filter with warping coefficient $\alpha=0.45$. This leads to $60$ log Mel-Frequency Spectral Coefficients (MFSCs), representing the harmonic features. For aperiodic features, we use WORLD’s inherently bandwise aperiodic analysis to reduce the dimensionality to $4$.


All features, both in the input and the target were normalized globally over the dataset to a range between $0$ and $1$ using min-max normalization across the dataset.



\subsection{Network Hyperparameters}
For training, samples of $N = 128$ consecutive time frames were randomly selected from the training set instead of sequentially feeding the network. We used minibatch training, with a batch size of $30$ batches per iteration. We used $k=5$ blocks of gated convolutions and $C=128$ filter channels for each of the convolutional layers except the final layer. We trained the network for $50k$ iterations.




.

\section{Results}
We compare our proposed system, henceforth referred to as \textit{sssynth}, to two open source source separation systems, DeepConvSep \cite{chandna2017monoaural} and FASST \cite{salaun2014flexible, ozerov2012general}. DeepConvSep is a source separation system based on CNNs, which uses the network to estimate time frequency masks. We chose this particular system, because it has been trained on the iKala dataset, that we are using for evaluation. The system was among the best algorithms in the MIREX2016 source separation challenge, which also used this dataset. It was also shown to be competitive with state-of-the-art algorithms for separating multiple sources from a musical mixture in the SiSEC 2016 challenge. The FASST algorithm is an open source implementation of an NMF based approach to source separation and has been used as a benchmark for evaluating many source separation algorithms over the years. 

\subsection{Objective Evaluation}
\begin{figure}[H]
\centering
\includegraphics[height = 3.7cm, width=0.5\textwidth]{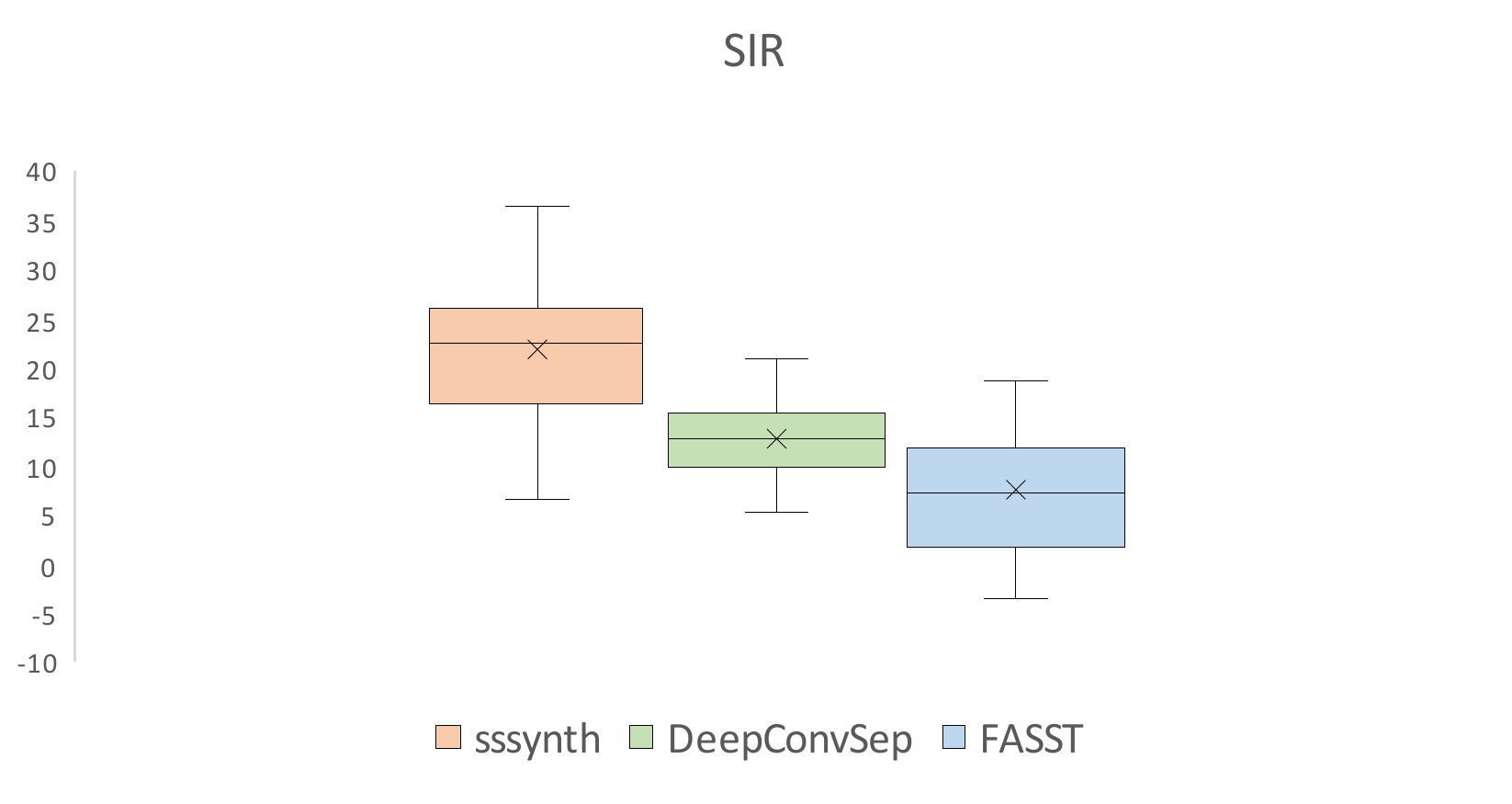}
 \caption{The SIR metric from the BSS Eval toolkit for the three systems to be compared. It can be observed that sssynth achieves a higher score in this metric than the other two systems. This is expected since the use of voice specific vocoder features in our system prevents interference from other sources in the output.}
 \label{fig:SIR}
\end{figure}

\begin{figure}[H]
\centering
\includegraphics[height=3.7cm, width=0.45\textwidth]{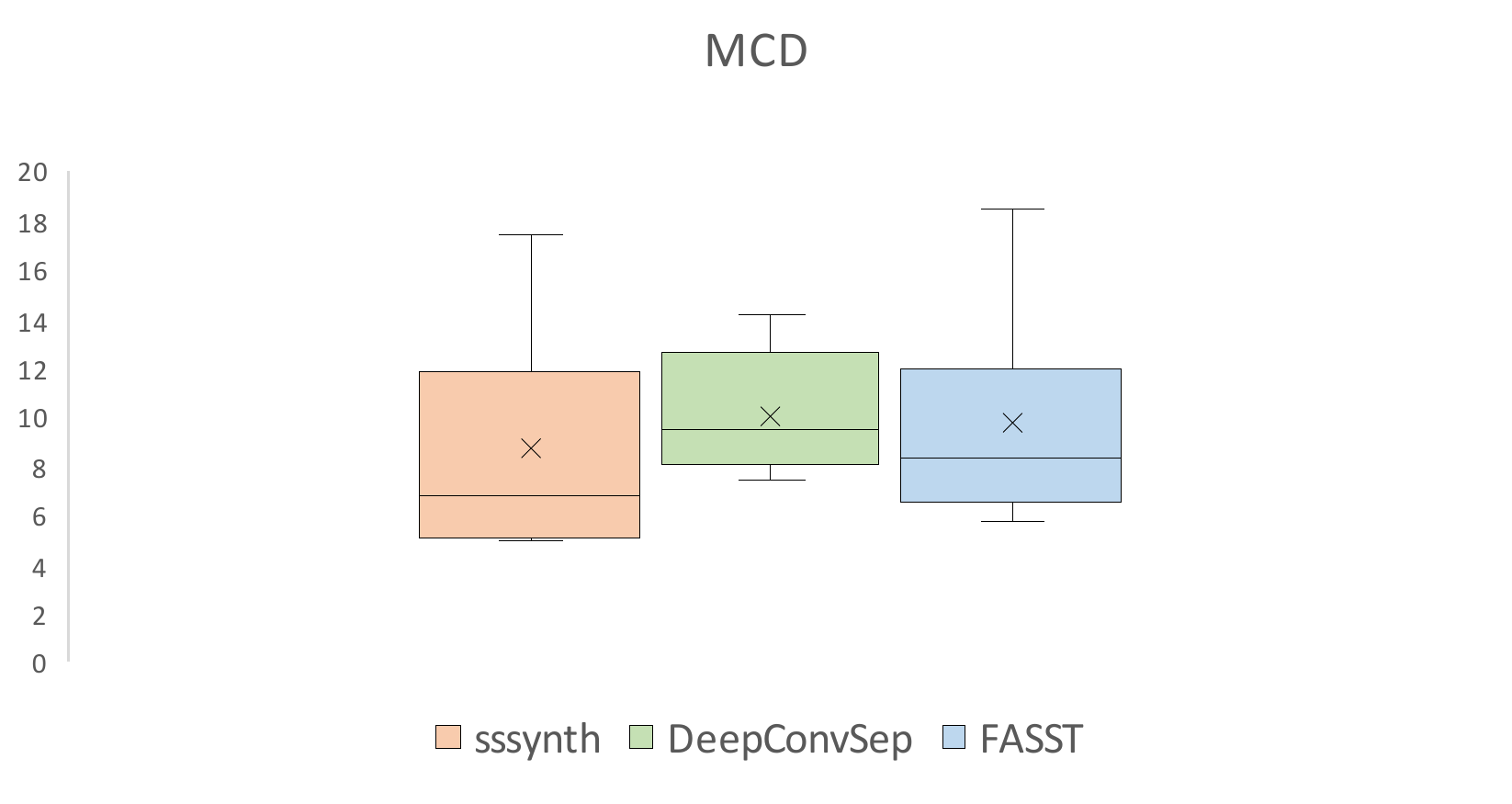}
 \caption{The Mel Cepstral Distortion (MCD), in dB, for the three systems to be compared. While the upper range of the MCD for the three algorithms is comparable, our system has a better lower range and mean than the other two to which is being compared.}
 \label{fig:cep_dist}
\end{figure}

We assess the quality of our system on three aspects; the intelligibility of the separated voice signal, the presence of unwanted sources in the signal and the auditory quality of the synthesized signal. 
We use Mel Cepstral Distrotion (MCD) as an objective measure for overall audio quality of the vocal synthesis and the Source to Interferences Ratio (SIR) from the BSS Eval\footnote{Other objective measures from the BSS Eval toolkit like SDR and SAR can be found at \url{https://pc2752.github.io/singing_voice_sep/}. We do not discuss them here as the results from these metrics do not correlate with our subjective evaluation. We believe that this is primarily because the output of our system is a re-synthesized version of the original signal and thus these measures do not give a fair reflection of the quality of the output for our case.} \cite{vincent2006performance} set of metrics for the amount of interference from other source in the output. The evaluation using these two is shown in Figures \ref{fig:SIR} and \ref{fig:cep_dist} respectively.  Intelligibility, however, is a subjective matter that we assess through a subjective listening test, along with the other two aspects.


\subsection{Subjective Evaluation}
For subjective evaluation, we presented the three evaluation criterion to listeners in the form of an online AB preference test, wherein the listener was asked to choose between two corresponding samples, given a predefined criteria. 

The three systems to be evaluated, sssynth, DeepConvSep and FASST, were paired for the test, resulting in $3$ pairs. Each of the criteria, Intelligibility, Interference and Audio Quality had $5$ samples from each of the $3$ pairs, resulting in $45$ separate questions. For questions related to Interference, the listener was presented with a reference mixture of the voice and backing track and was asked to pick the system which had less residual from the backing track present with the vocal track. For audio quality related questions, the listener was presented with a reference of the clean vocals and was asked to choose the system which was closer to the reference in terms of audio quality. Finally, for the questions related to intelligibility, the listener was asked to choose the system which was more easily intelligible, without a reference audio, which might have caused a bias.  

\begin{figure}[H]
\centering
\includegraphics[width=0.45\textwidth]{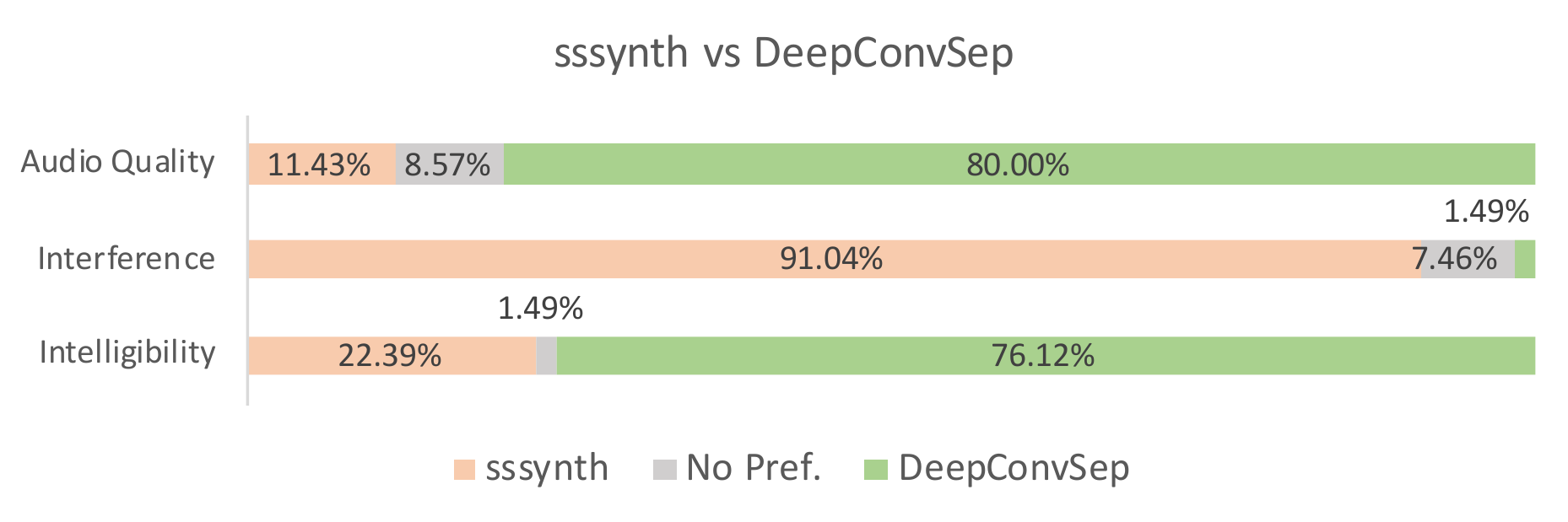}
 \caption{Results of the subjective evaluation comparing sssynth to DeepConvSep. It can be seen that while DeepConvSep is preferred over our proposed model for the intelligibility and audio quality criterion, sssynth is perceived to perform better in terms of interference, by a majority of the people participating in the listening test.}
 \label{fig:sub_int}
\end{figure}

\begin{figure}[H]
\centering
\includegraphics[width=0.45\textwidth]{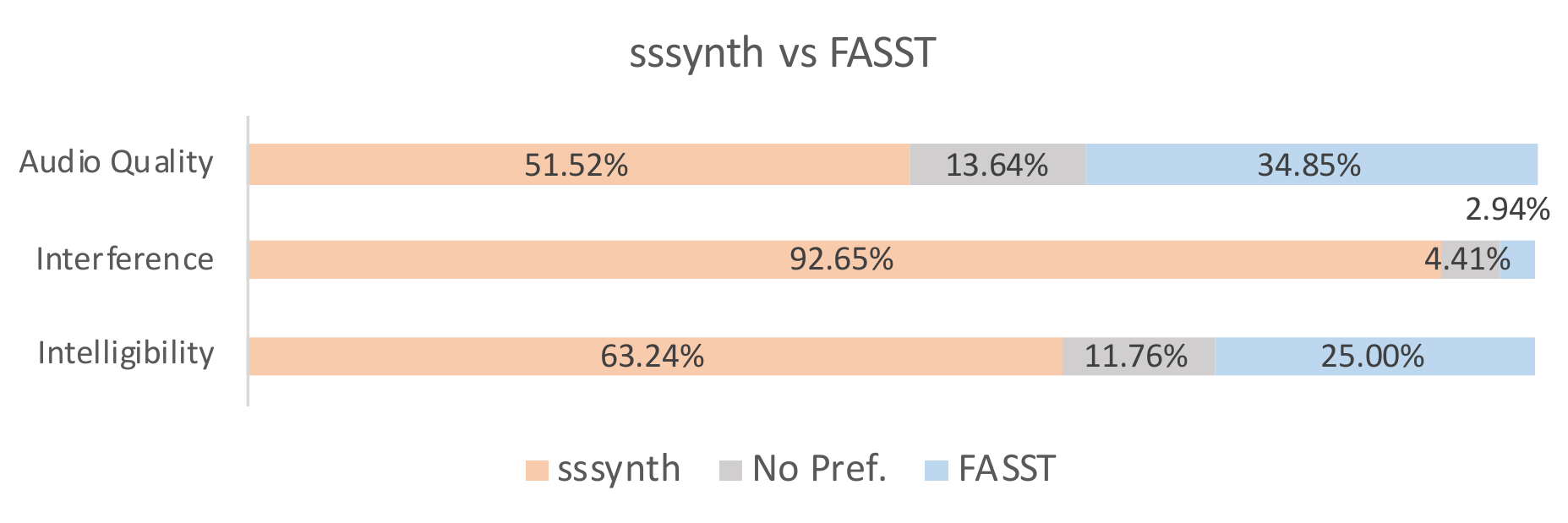}
 \caption{Results of the subjective evaluation comparing sssynth to FASST. While a clear preference towards sssynth is observed for the intelligibility and interference criterion, the results are more evenly divided for the case of audio quality.}
 \label{fig:sub_sep}
\end{figure}

\begin{figure}[H]
\centering
\includegraphics[width=0.45\textwidth]{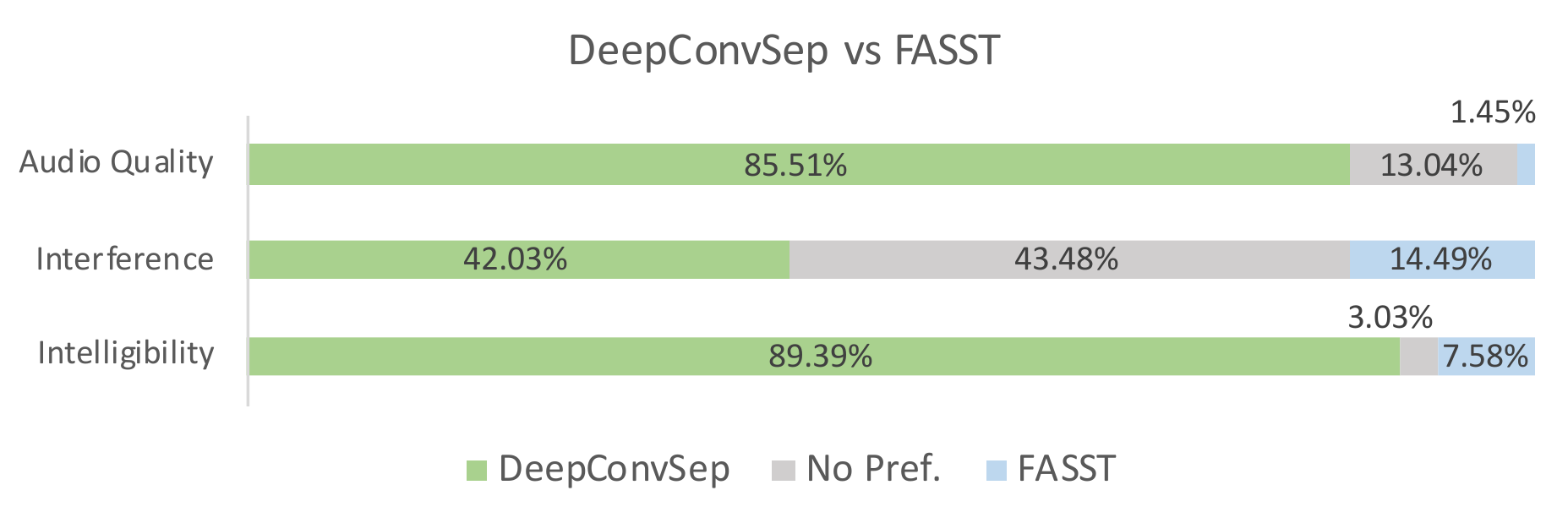}
 \caption{Results of the subjective evaluation comparing DeepConvSep to FASST. The participants in the listening test seem to show a clear preference towards DeepConvSep for the audio quality and intelligibility citerion. However, for the case of interference, no clear preference is observed.}
 \label{fig:sub_qual}
\end{figure}

We used $5$ second samples from songs in the test set, not used for training the model. Since the song examples in the iKala dataset are primarily in the Mandarin Chinese language, the online listening test, was presented in the Mandarin Chinese language\footnote{\url{http://mtg.upf.edu/sourcesepeval/}}. $16$ participants, fluent in the Mandarin Chinese language participated in the survey and the results are presented in Figures \ref{fig:sub_int}, \ref{fig:sub_sep} and \ref{fig:sub_qual}.  

It can be seen from Figure \ref{fig:SIR}, that the presented system, sssynth preforms better than DeepConvSep and FASST in terms of the SIR metric. This metric is a measure of the amount of interference from background sources in the separated source and the observation agrees with the results of the subjective evaluation shown in Figures \ref{fig:sub_int} and \ref{fig:sub_sep}. This is expected since the use of the vocal specific vocoder features limits leakage from other sources into the estimated vocal track. 

For audio quality, while the objective evaluation, shown in Figure \ref{fig:cep_dist}, seems to favour sssynth, the subjective listening test (Figure \ref{fig:sub_int}) shows that the audio quality and the related aspect of intelligibility is better for DeepConvSep. We believe that the interference from the backing track in DeepConvSep leads to higher energy in some mel bands thus leading to a higher distortion factor. However, since the output of the DeepConvSep algorithm is basically a synthesis of a masked version of the mixture magnitude spectrogram, using the phase of the 
mixture, the perceived audio quality is similar to that of the original signal. Our model is based on the WORLD vocoder parameters and thus avoids interference from background signals, leading to a lower distortion factor, However, since it a re-synthesis of the voice signal, the perceived quality is not as high as that one can achieve from the masking technique. The relative lack of audio frames with consonant sounds, compared to vowel sounds, also influences intelligibility.
However, our model outperforms the FASST NMF based algorithm in both the intelligibility and audio quality criterion. 

\section{Conclusions And Future Work}
We have proposed and evaluated a vocoder based voice extraction algorithm for the case of musical signals. From subjective and objective evaluation, we find that while the vocoder does a good job of isolating the voice signal from the background music, it falls behind state-of-the-art source separation algorithms in terms of audio quality. Concurrently, the intelligibility of our system also leaves some room for improvement. We are currently working on integrating a WaveNet based vocoder system which can alleviate these problems. Also, we will implement autoregression in our model, which we believe will improve audio quality with a compromise in inference time. We also plan to use the vocoder features output by our system directly for other tasks such as lyrics extraction, score alignment and voice transformation. 

\section{Acknowledgements}
\label{sec:acknowledge}

The TITANX used for this research was donated by the NVIDIA Corporation. This work is partially supported by the Towards Richer Online Music Public-domain Archives (TROMPA) project.

\vfill\pagebreak


\bibliographystyle{IEEEbib}

\bibliography{mybib}
\end{document}